\documentstyle[ltwol]{article}

\begin{document}

\title{YANG--MILLS DUALITY AS ORIGIN OF GENERATIONS, QUARK MIXING, AND
NEUTRINO OSCILLATIONS}

\author{TSOU SHEUNG TSUN}

\address{Mathematical Institute, Oxford University\\
24--29 St.\ Giles', Oxford OX1 3LB,
United Kingdom\\E-mail: tsou\,@\,maths.ox.ac.uk}   

\twocolumn[\maketitle\abstracts{
The origin of fermion generations is one of the great mysteries in particle
physics.  We consider here a possible solution within the Standard Model 
framework based on a nonabelian generalization of electric-magnetic duality.  
First, nonabelian duality says that dual to the colour (electric) symmetry 
$SU(3)$, there is a ``colour magnetic symmetry'' $\widetilde{SU}(3)$, which 
by a result of 't~Hooft is spontaneously broken and can thus play the role 
of the "horizontal symmetry" of generations.  Second, nonabelian
duality 
suggests the manner this symmetry is broken with frame vectors in internal 
symmetry space acting as Higgs fields.  As a result, mass matrices 
factorize leading to fermion mass hierarchy.  At the tree level, there is 
no mixing but with loop corrections, the mass matrices rotate and mixing 
occurs.  A calculation to first order gives mixing (CKM and MNS)
matrices 
in general 
agreement with experiment.  In particular, quark mixing is seen naturally 
to be weak compared with leptons, while within the lepton sector, $\mu-\tau$ 
mixing turns out near maximal but $e-\tau$ mixing small, just as seen in 
recent $\nu$ oscillation experiments.  In addition, the scheme leads to 
many testable predictions ranging from rare FCNC meson decays and $\mu-e$ 
conversion in nuclei to cosmic ray air showers above $10^{20}$ eV,
which will be detailed in the followng talk by Chan.
}]

\section{Yang--Mills Duality}
In electromagnetism, the dual field tensor ${}^*\!F_{\mu\nu}$ is given
in terms of the field tensor $F_{\mu\nu}$ by a duality transformation
which is just the Hodge star operation:
\begin{equation}
{}^*\!F_{\mu\nu} = - \textstyle{{1}\over{2}}
\epsilon_{\mu\nu\rho\sigma}
F^{\rho\sigma}.
\label{hodge}
\end{equation}

Under this operation:
\begin{eqnarray*}
\mathbf{E} & \longleftrightarrow & \mathbf{B} \\
{\rm electric\ charge} & \longleftrightarrow & {\rm magnetic\
monopole} \\
e & \longleftrightarrow & \tilde{e}
\end{eqnarray*}
where $e \tilde{e} = 2 \pi$.  As is well-known, electromagnetism is
symmetric with respect to this duality transform.

A natural question to ask is whether the symmetry persists in
nonabelian Yang--Mills theory.  If we again use the Hodge star as our
duality transform, then the question was answered in the negative by
Gu and Yang~\cite{guyang}, who gave counter-examples where
$F_{\mu\nu}$ satisfies the source-free Yang--Mills equation
but ${}^*\!F_{\mu\nu}$, although defined, is not a gauge field.

So if we want to generalize electric--magnetic duality to the
nonabelian case, then we have to look for a generalization
of the Hodge star.  Preferably this generalized dual transform
$(\tilde{\ })$ should satisfy the following properties:
\begin{enumerate}
\item $(\quad)^{\sim\sim} = \pm (\quad)$,
\item electric field $F_{\mu\nu} \stackrel{\sim}{\longleftrightarrow}
$ magnetic fields $\tilde{F}_{\mu\nu}$,
\item both $A_\mu$ and $\tilde{A}_\mu$ exist as potentials (away from
charges and monopoles),
\item magnetic charges are monopoles of $A_\mu$, and electric charges
are monopoles of $\tilde{A}_\mu$,
\item $\tilde{\ }$ reduces to ${}^*$ in the abelian case.
\end{enumerate}

A hint of what we should be looking for comes from a result of Wu and
Yang~\cite{wyphase}, namely that what describes gauge theory {\em
exactly} is neither the gauge potential $A_\mu$ nor the gauge field
$F_{\mu\nu}$, but the Dirac phase factor or Wilson loop
\begin{equation}
\Phi (C)=P \exp ig \int_C A_\mu (x) dx^\mu,
\end{equation}
where $P$ denotes path ordering with respect to the loop $C$.  This
leads to a loop space formulation of Yang--Mills 
theory~\cite{gaugebook,feynloop}, 
as originally
proposed by Polyakov~\cite{polyakov}.  And it is in terms of these loop
variables that we were able to formulate a generalized 
transform~\cite{prenabdual}
satisfying the above 5 conditions, and to show that 
Yang--Mills theory is
completely dual symmetric in this sense~\cite{ymduality,dsmrth98}.

What is relevant for what follows is that as a result of this dual
symmetry the gauge symmetry is doubled, so that if $F_{\mu\nu}$ is the
gauge field of an $SU(N)$ Yang-Mills theory, then the total symmetry
is $SU(N) \times \widetilde{SU(N)}$, where $\widetilde{SU(N)}$ is
abstractly the same group $SU(N)$ but has the opposite 
parity~\cite{ymduality}.

\section{Generation Puzzle}
Broadly speaking, the generation puzzle consists of three questions
with as yet no generally accepted answers:
\begin{itemize}
\item Why are there 3 and only 3 generations?
\item Why is there mass hierarchy?
\item Why is there nontrivial mixing?
\end{itemize}

The 3 generations are: for the up-quarks $t,c,u$; for the 
down-quarks $b,s,c$; for the neutrinos
$\nu_\tau,\nu_\mu,\nu_e$; and for the charged leptons $\tau,\mu,e$.
As far as we can observe, these triples differ only in their masses,
but then to such an extent that they are `hierarchical'.  Thus:
$$\begin{array}{lll}
m_t \sim 180\ {\rm GeV}\ \ & m_c \sim 1.1\ {\rm GeV}\ \ & m_u \sim 4\ {\rm
MeV} \\
m_b \sim 4.2\ {\rm GeV}\ \ & m_s \sim 120\ {\rm MeV}\ \ & m_d \sim 7 \ {\rm
MeV}  \\
m_\tau \sim 1.8\ {\rm GeV}\ \ & m_\mu \sim 100\ {\rm MeV}\ \ & m_e \sim 0.5\
{\rm MeV} 
\end{array}$$

By mixing we mean that up and down states are not aligned but related
by the empirical matrices~\cite{databook} (absolute values only), for 
quarks and leptons respectively:
\begin{equation}
|V_{\rm CKM}| = \left( \begin{array}{lll} 0.975\ & 0.222\ & 0.003 \\
                                    0.222\ & 0.974\ & 0.040  \\
                                    0.009\ & 0.039\  & 0.999
\end{array} 
\right),
\label{ckmmat}
\end{equation}
\begin{equation}
|U_{\rm MNS}| = \left( \begin{array}{ccc} ?\ \ & 0.4 - 0.7 & 0.0 - 0.15 \\
                                    ?\ \ & ? & 0.45 - 0.85 \\
                                    ?\ \ & ? & ? \end{array} \right),
\label{mnsmat}
\end{equation}
with the following three noticeable features:
\begin{itemize}
\item mixing is smaller for quarks and larger for leptons,
\item the corner elements in both are very much smaller,
\item the (23) element for leptons is much larger than for quarks.
\end{itemize}

\section{Dualized Standard Model}
The Dualized Standard Model~\cite{physcons,dsmrph98,dualgen} (DSM)
tries to answer all these questions by
making use of the dual symmetry we proved (\S 1) and applying it to the
Standard Model.  One immediate consequence is that the gauge group is
now doubled to
$$ SU(3) \times SU(2) \times U(1) \times \widetilde{SU(3)} \times
\widetilde{SU(2)} \times \widetilde{U(1)}. $$

In the application we are concerned with in this article, we need only
focus on the colour and dual colour symmetries, that is: $SU(3) \times
\widetilde{SU(3)}$.  Now by a theorem of 't~Hooft~\cite{thooft}, 
we know that the
fact that colour $SU(3)$ is confined implies that dual colour 
$\widetilde{SU(3)}$ is broken~\cite{comrel}.  We now make the crucial
assumption that this broken $\widetilde{SU(3)}$ is
the generation symmetry, in other words, a very specific type of
horizontal symmetry~\cite{horizontal}.  The only other assumption that 
DSM makes is about
the role of Higgs fields as frame vectors.  For want of space, we
shall not elaborate on this.

As a reuslt, the fermion mass matrix at tree level is of rank 1:
\begin{equation}
m = m_T \left( \begin{array}{c} x \\ y \\ z \end{array} \right)
      (x, y, z),
\label{mmat}
\end{equation}
where $(x,y,z)$ is a normalized vector and 
$m_T$ is essentially the mass of the heaviest generation.  With
only one nonzero eigenvalue we already have a good tree-level
approximation to the mass hierarchy.
Under
renormalization (to one loop) $m$ remains factorized, but 
\begin{equation}
\frac{d}{d\mu} \left( \begin{array}{c} x \\ y \\ z \end{array} \right)
   = \frac{5}{32 \pi^2} \rho^2 \left( \begin{array}{c} x_1 \\ y_1 \\ z_1
      \end{array} \right),
\label{dsmrge}
\end{equation}
where $\rho$ is a (fitted) constant and
\begin{equation}
x_1 = \frac{x(x^2-y^2)}{x^2+y^2} + \frac{x(x^2-z^2)}{x^2+z^2},
   \ \ \ {\rm cyclic}.
\end{equation}

So we see that the vector $(x,y,z)$ rotates with the
energy scale where the rotation depends on the fermion-type, so that up
and down states become disoriented with respect to each other leading 
to nontrivial mixing matrices.  At the same time, mass starts to 
``leak'' from the top generation into the two lower generations giving
them small but nonzero masses.   As the energy 
changes,
the vector $(x, y, z)$ traces out a trajectory on the unit 
sphere, starting at high energy from near the fixed point $(1,0,0)$ and 
moving towards the low-energy fixed point $\frac{1}{\sqrt{3}}(1,1,1)$.
Although the trajectories can in principle be different for different
fermion-types, the data demand, for reasons yet theoretically unclear, that 
they coincide to a very good approximation.  The 12 different fermion states
thus occupy only different points on 
this single trajectory (Figure 1 of reference~\cite{features}), 
from which we are already
able to deduce, qualitatively and even quantitatively, the three
features of mixing listed above~\cite{features}.

This follows from elementary differential geometry.  For a given fermion type,
the state vector for the
top (i.e.\ heaviest) generation is the radial vector $(x, y, z)$ at
the appropriate scale, 
and the state vector
for the second generation is approximately the tangent vector to both
curve and surface.  These together with the state vector for the 
lightest generation form the Darboux
triad~\cite{docarmo} at the
position of the top generation.  So we have two such triads at the
positions of the up and down fermions, and the direction cosines between
them give the corresponding mixing matrix.  To first order in the
separation $\Delta s$ of the two top fermion positions (i.e.\ at $t$
and $b$ for quarks and $\nu_3$ and $\tau$ for leptons), this is given
by the Serret-Frenet-Darboux formulae as
\begin{equation}
\left( \begin{array}{ccc}
   1 & - \kappa_g \Delta s & - \tau_g \Delta s \\
   \kappa_g \Delta s & 1 & \kappa_n \Delta s \\
   \tau_g \Delta s & - \kappa_n \Delta s & 1  \end{array} \right),
\label{Muimat}
\end{equation}
with $\kappa_n$ the normal curvature, $\kappa_g$ the geodesic curvature,
and $\tau_g$ the geodesic torsion of a curve on a surface.  For the unit
sphere, $\kappa_n = 1$ and $\tau_g = 0$.  From this we deduce first that the
corner elements are of second order in $\Delta s$ and therefore
small compared with the others.  Secondly,
we conclude that the 23 and 32 elements are given approximately just by
the separation $\Delta s$.  And indeed, if 
one takes the trouble to measure with a bit of string these distances on
the the trajectory in Figure 1 of reference~\cite{features}, 
one will find values very
close to the experimental numbers given for these elements in (\ref{ckmmat})
and (\ref{mnsmat}).

Of course, having actually done the calculation~\cite{ckm,nuosc,phenodsm}, 
one can make much a more
detailed comparison with data than the above purely geometric
picture.  One finds that all the mixing parameters overlap with the present
experimental limits, except for the solar neutrino mixing element $U_{e2}$,
which being related to the trajectory-dependent geodesic curvature according 
to (\ref{Muimat}) is particular difficult for our calculation to get correct.
We note that all these numbers have been obtained by adjusting only one 
parameter to the Cabibbo angle $V_{us} \sim V_{cd}$, the other two parameters 
in the calculation having already been fitted to fermion 
masses~\cite{phenodsm}.

\section{Further Results}
By making the assumption that dual colour, which exists by virtue of
Yang--Mills duality, is generation symmetry, DSM is able to account for
the origin of exactly 3 generations with generally correct mixing.  In
addition, as reported in the same volume~\cite{condugen}, the scheme
leads to many testable predictions, including rare FCNC meson
decays~\cite{fcnc}, $\mu - e$ conversion in nuclei~\cite{mueconv}, lepton
transmutational effects~\cite{impromat,photrans}, and even cosmic ray
air showers above the GZK bound~\cite{airshow2}.

\end{document}